\documentstyle[11pt,aaspp4]{article}

\def\lesssim{\, 
\lower2truept\hbox{${<\atop\hbox{\raise4truept\hbox{$\sim$}}}$}\,}
\def\gtrsim{\, 
\lower2truept\hbox{${> \atop\hbox{\raise4truept\hbox{$\sim$}}}$}\,}

\accepted{2 April 1997}
\journalid{486}{September 1997}

\slugcomment{To appear in The Astrophysical Journal, September 1997}

\begin{document}

\title{Thick tori around AGN: the case for extended
tori and consequences for their X--ray and IR emission}

\author{Gian Luigi\ Granato\altaffilmark{1}}
\affil{Osservatorio Astronomico di Padova, Padova, Italy}

\author{Luigi\ Danese\altaffilmark{2}}
\affil{International School for Advanced Studies, SISSA, Trieste, Italy}

\and

\author{Alberto Franceschini\altaffilmark{3}}
\affil{Dipartimento di Astronomia di Padova, Padova, Italy}

\altaffiltext{1}{Osservatorio Astronomico, Vicolo dell'Osservatorio 5,
I35122 Padova, Italy. E-mail: Granato@pdmida.pd.astro.it\\
Tel. (049) 8293441, Fax (049) 8759840}
\altaffiltext{2}{International School for Advanced Studies, Strada
Costiera
11, I34014 Trieste, Italy. E-mail: Danese@pdmida.pd.astro.it}
\altaffiltext{3}{Dipartimento di Astronomia, Vicolo Osservatorio 5, I-35122,
Padova, Italy. E-mail: Franceschini@pdmida.pd.astro.it}

\begin{abstract}
Two families of models of dusty tori in AGNs (moderately thick
and extended versus very thick and compact) are tested against
available observations. The confrontation suggests that the former class
better explains the IR broad--band spectra of both broad and narrow
line AGNs,  the anisotropy of the emission deduced by comparing
IR properties of Seyfert 1 and 2 nuclei, the results of 
IR spectroscopy  and those of high spatial resolution observations. 
There is however clear evidence for a broad distribution of 
optical depths. We also examine the
relationship between IR and X--ray emission. The data support
a view in which the matter responsible for the X--ray absorption is mostly
dust free, lying inside the dust sublimation radius.  The consequences of
these results for the hard X--ray background as well as IR counts and
background are discussed.
\end{abstract}

\keywords{galaxies: active -- galaxies: nuclei --
galaxies: Seyfert -- infrared: galaxies - radiative transfer
ISM: dust, extinction}

\section{Introduction}

Direct and indirect evidence of obscuring material around Active
Galactic Nuclei (AGN) has been accumulated. High spatial resolution
observations of emission from molecules and dust show that enriched gas
is abundant around AGN on scales ranging from a fraction of parsec (see
e.g. Miyoshi et al.\ 1995; Greenhill et al.\ 1995) to many tens of
parsec (see e.g.  Braatz et al.\ 1993; Jackson et al.\ 1993; Cameron et
al.\ 1993; Tacconi et al.\ 1994; Genzel et al.\ 1995).
\nocite{MMH95,GJM95,BWG93,JPI93,CSS93,TGB94,GWT95}

These results strongly support unified schemes of AGN, according to
which the distinction of AGN in  broad and narrow line objects is mainly
due to orientation of circumnuclear dust structures. These schemes have
been originally invoked to explain the results of  spectropolarimetric
observations, namely that NGC 1068 (the prototype of Seyfert 2 galaxies)
and several additional narrow line AGN show broad line components in
their spectra (see Antonucci 1993 for a comprehensive review). 
\nocite{Ant93}  Direct 
view of the Broad Line Region (BLR) is blocked by an optically thick
structure. The broad lines detected with spectropolarimetry are the
result of the scattering into the line of sight of the radiation from
the BLR by free electrons and/or dust particles above the nucleus.
Moreover, in some optically narrow line AGN, broad lines have been
detected with  spectroscopy in the near--IR, where the absorption 
is less effective
than in the optical bands (Rix et al. 1990; Ruiz, Rieke
and Schmidt 1994; Goodrich, Veilleux and Hill 1994). 
\nocite{RRR90,RRS94,GVH94}

X--ray spectra of narrow line AGN exhibit absorption significantly
larger than type 1 AGN, providing strong support for the obscuring
hypothesis (see e.g.\ Nandra and Pounds 1994; Smith and
Done 1996). \nocite{NaP94,SmD96}

The anisotropy of the nuclear radiation (expected to be collimated in a
cone with apex in the nucleus and opening angle determined by the
absorbing structure) has been confirmed by images of emission lines of
[O III], [N II], $H_{\alpha}$ and other elements for a significant
number of narrow line AGN (see Wilson 1996 for a review ) which
show conical regions of high excitation emission--line extending from
the active nuclei. \nocite{Wil96}

More recently, the presence of obscuring tori has been also invoked to
explain the difference between narrow-- and broad--line radiogalaxies
(Barthel 1989). In this context the failed detection of CO
J=1--0 absorption from Cygnus A was embarrassing (Barvainis and
Antonucci 1994).  However there are explanations which allow
to keep the presence of a dusty and molecular torus. In particular,
Maloney, Begelman and Rees (1994) showed that the nonthermal
radio continuum may increase the excitation temperature of the lower
rotational level thus reducing the optical depth.  
Therefore the lack of CO absorption
may be a general property of the radio galaxies. Also
photo--dissociation, ionization and heating by X-ray tend to decrease
the CO absorption optical depth.  
\nocite{Bar89,BaA94,MBR94}

The energy absorbed by the obscuring structure must be reradiated in the
infrared. IR broad--band spectra of nuclei of Seyfert galaxies can be
fitted by models in which a significant fraction of the nuclear
emission is reprocessed by  an axisymmetric torus--like structure (Pier
\& Krolik 1992, 1993; Granato \& Danese 1994; 
Efstathiou \& Rowan-Robinson 1994). On the contrary non thermal 
emission mechanisms are not able to
reproduce the extremely steep decline of the continuum observed
between 100 and 1000 $\mu$m (Hughes et al.\ 1993).
\nocite{PiK93,PiK92,GrD94,EfR94,HRD93}

X--ray emission may also be affected by the presence of dusty molecular
tori around the active nuclei. Madau, Ghisellini \& Fabian (1993) and
Krolik, Madau \& Zycki (1994) have computed X--ray spectra emerging from
nuclear region along directions with significant Thomson optical depths.
Scattering and absorption significantly modify intrinsic spectra,
yielding spectral shapes suitable to reproduce the 2--100 keV X--ray
Background as integrated emission from partially covered AGN (Madau et
al.\ 1993; Comastri et al.\ 1995). \nocite{MGF93,KMZ94,CSZ95}

The absorbing structures around the active nuclei have usually been
modelled as axial symmetric tori (Pier \& Krolik 1992, 1993; Granato \&
Danese 1994; Efstathiou \& Rowan--Robinson 1994), though different
geometries, like warped discs, have also been suggested (Sanders et al.\
1989). Models proposed by Pier \& Krolik (1992, 1993) are characterized
by extremely large optical thickness ($\tau \gtrsim 1000$ in the UV,
which entails  $\tau \gtrsim 10$ at 10$\mu$m and Thomson optical depth
$\tau_T\gtrsim 1$) and compactness (radial dimension $\lesssim$ a few
pc).  By converse models proposed by Granato \& Danese (1994) are based
on tori with optical depths in the UV band ranging from 10 to 300 and 
with maximum radii ranging from tens to hundreds pc.
\nocite{SPN89,PiK93,PiK92,GrD94,EfR94}

In Section \ref{secextcom} the two families of models (moderately thick
and extended versus very thick and compact) are tested against
observational information: IR broad--band spectra (Sect.\
\ref{subsecsed}) and its anisotropy (Sect.\ \ref{subsecani}), IR
spectroscopy (Sect.\ \ref{subsecspe}) and high spatial resolution
observations (Sect.\ \ref{subsechsr}).  In Section \ref{secxir} 
we examine the
links between IR and X--ray emission and we show that the data support
the view that the matter responsible for most of
the X--ray absorption is dust
free and lies inside the dust sublimation radius.  The consequences of
these results for the hard X--ray background as well as IR counts and
background are then discussed in Section \ref{secstat}), where we estimate
in particular the dusty AGN counts in the IR bands relevant to ISO. The
final Section is devoted to summarize our conclusions.

A Hubble constant $H_0= 50 \, \rm{km} \, \rm{s}^{-1} \, \rm{Mpc}^{-1}$
and a deceleration parameter $q_0=1/2$ have been assumed whenever
necessary.

\section{Extended and compact tori and confrontation with the data}
\label{secextcom}

As mentioned above, extended models have been investigated by
Granato and Danese (1994) (GD hereafter) and by Efstathiou and
Rowan--Robinson (1994), while compact and extremely thick models 
($\tau_T\gtrsim 1$ and correspondengly $A_V\gtrsim 800$) have
been investigated by Pier and Krolik (1992, 1993) (PK hereafter).

GD discussed a numerical code which solves the radiative
transfer equation in a circumnuclear dust distribution.  This 
step is required since in the tori predicted by unified models the dust
emission is self--absorbed at least in the near-- and mid--IR.  The code
is quite flexible concerning the geometry and composition of the dusty
medium, the only restriction being axial simmetry, and thus allows
a wide exploration of the parameter space, including both classes of
models under discussion.

In the computations a standard galactic dust composition is assumed.
The inner radius $r_{in}$ of the dust distribution is set by the grain
sublimation condition, above $T_s=1500$ for graphite and $T_s=1000$ for
silicates. This translates into $r_{in}\sim 0.5 L_{46}^{1/2}$ pc, where
$L_{46}$ is the luminosity of the primary optical--UV emitter in units
of $10^{46}$ erg $s^{-1}$.

\begin{figure*}
\epsscale{1.0}
\plotone{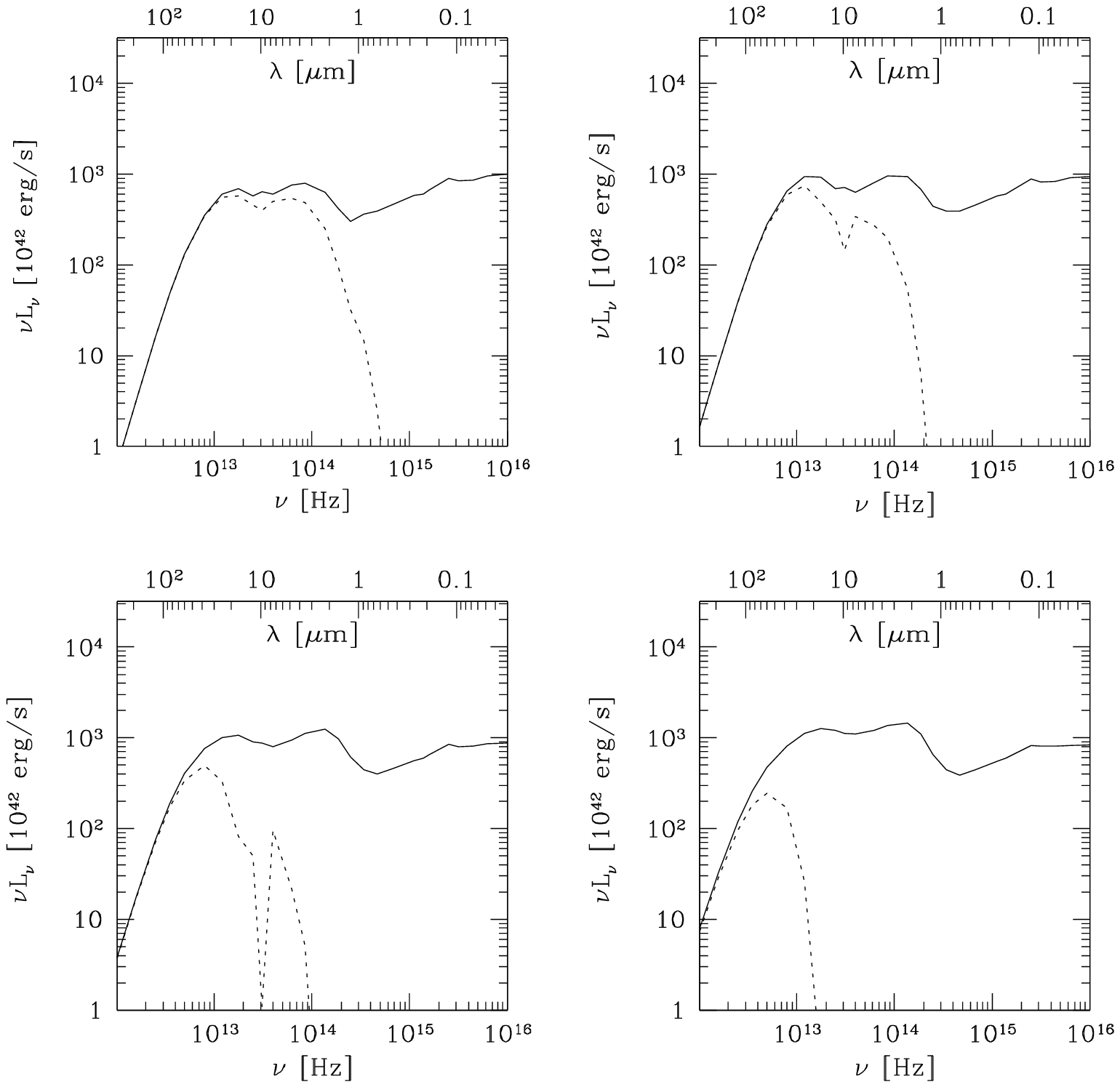}
\caption{Examples of SEDs predicted by our radiative transfer code.
In these cases the torus is homogeneous, has a covering factor of 0.8
and $r_{out}/r_{in}$ is set to 300. From top to bottom and from left to 
right the adopted optical thickness along obscured directions are 
$A_V = 10$,  30, 100 and 300. In each panel the solid line refer 
to a polar line of sight whilst the dashed line is the SED observed 
from the equator.}
\label{figexa}
\end{figure*}

The details of the model, as well as the effect of various free
parameters, have been already discussed by GD.  In this paper we wish to
focus mainly on the radial extension of the dust distribution, measured
by the ratio between the outer and inner radii $r_{out}/r_{in}$, and on
the absorption $A_V$ along typical obscured directions.  In
Fig.~\ref{figexa} we report representative SEDs predicted by our code
for a broad range of $A_V$.  The average
SED of the GD Seyfert 1 sample requires values for $r_{out}/r_{in}$ of
several hundreds and $A_V \sim 30$.  Roughly speaking, the first
parameter is related to the broadness of the IR bump arising from dust
reprocessing, whilst the latter controls mainly the near-IR 
slope of the SEDs as
observed from obscured directions as well as its anisotropy.  Therefore
$A_V$ is in principle constrained by the SEDs of obscured AGNs
(Sect.~\ref{subsecsed}) or by testing the anisotropy of mid--IR emission
(Sect.~\ref{subsecani})

\subsection{IR broad band spectra}
\label{subsecsed}

\begin{figure}
\epsscale{1.0}
\plotone{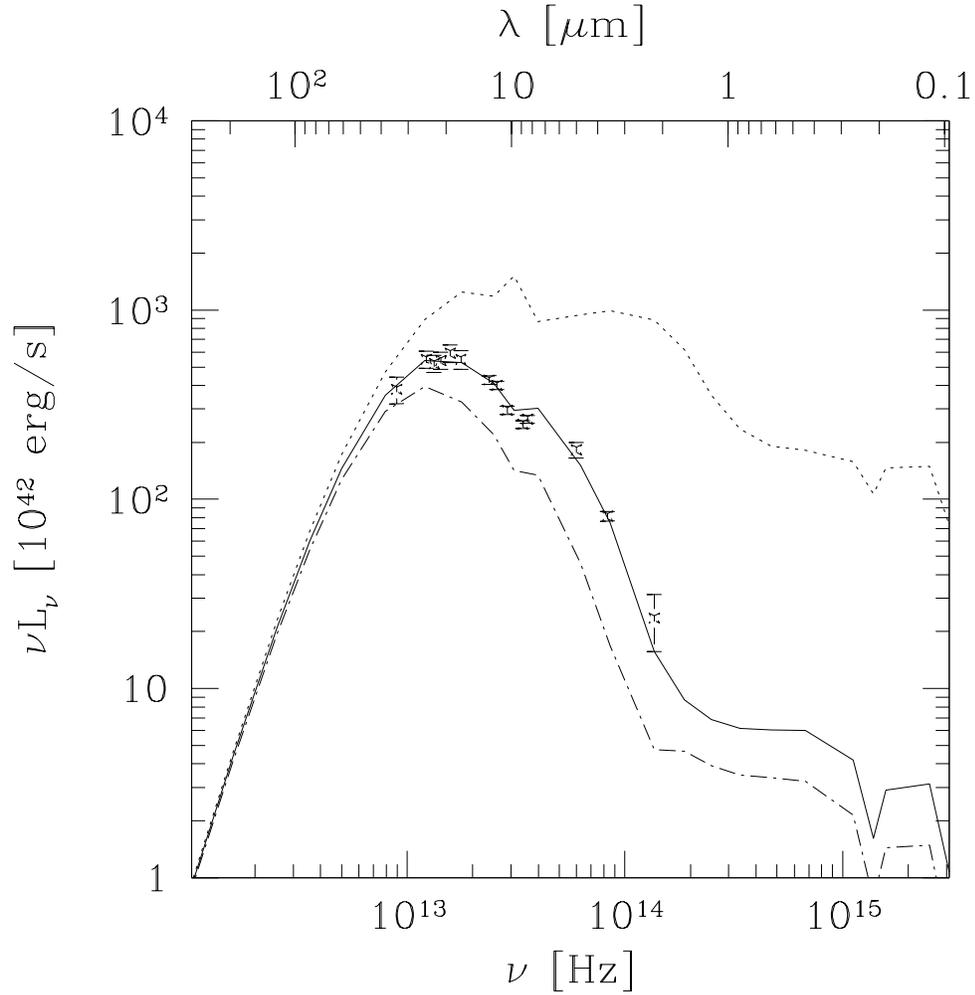}
\caption{
Fit to the observed IR SED of NGC 1068 (data from Rieke and Low 1975).
A distance of 22 Mpc has been assumed.  The dust density is independent
of $r$, but scales with the polar angle $\Theta$ as $\exp(-6\, \cos^2
\Theta)$.  The model is extended, having $r_{out}/r_{in} = 150$ and
$r_{out} \simeq 30$ pc, and it is characterized by  a `moderate' optical
thickness $A_V = 72$ along the adopted line of sight at $\Theta=65^o$
($A_V = 210$ at the equator).  The dotted and dashed lines are the
polar and equatorial SEDs respectively.}
\label{fig1068}
\end{figure}

In  Fig.~\ref{fig1068} we present a good fit to the overall IR
spectrum of NGC 1068, obtained with an extended and moderately thick
torus characterized by an optical thickness smoothly dependent on the
line of sight. In this model we used $A_V \simeq 72$ along the line of
sight, while the absorption is $A_V \simeq 220$ in the equatorial plane.
Assuming a distance $D=22$ Mpc, the required primary
optical--UV luminosity is $\simeq 1.5 \times 10^{45}$ erg/s, in nice 
agreement with the value obtained by UV spectropolarimetry with HST
(Antonucci et al.\ 1994). \nocite{AHM94} The
total dust mass involved is $2.7 \times 10^4 M_\odot$, consistent with
the estimates of gas masses ($\sim 10^7 M_\odot$) coming from CO
observations.  Though it is possible to
obtain similar fits with different combinations of the free parameters
and/or dust density laws, the extension and the optical depth,
particularly along the line of sight, are reasonably well constrained to
$r_{out}/r_{in}\simeq 100 - 200$ and $A_V \simeq 50-100$.

The information on IR nuclear spectra of Seyfert 2 galaxies other than
NGC 1068 remains rather scanty. Attempts of evaluating near IR nuclear
fluxes are hampered by low fluxes of the nuclei with respect to the host
galaxies.  Moreover, data on the IR spectra of the nuclear
regions have been collected through observations at various frequencies
with significantly different spatial resolution. As a consequence, we
must be cautious when referring to data collections as ``nuclear SEDs''.
Actually, only in a few cases of nearby Seyferts the data have angular
resolution good enough to allow an estimate of the nuclear fluxes.
Therefore here we restrict ourselves to objects within a distance of 50
Mpc, where an angle of 2 arcsec subtends about 0.5 Kpc.

\begin{figure}
\epsscale{1.0}
\plotone{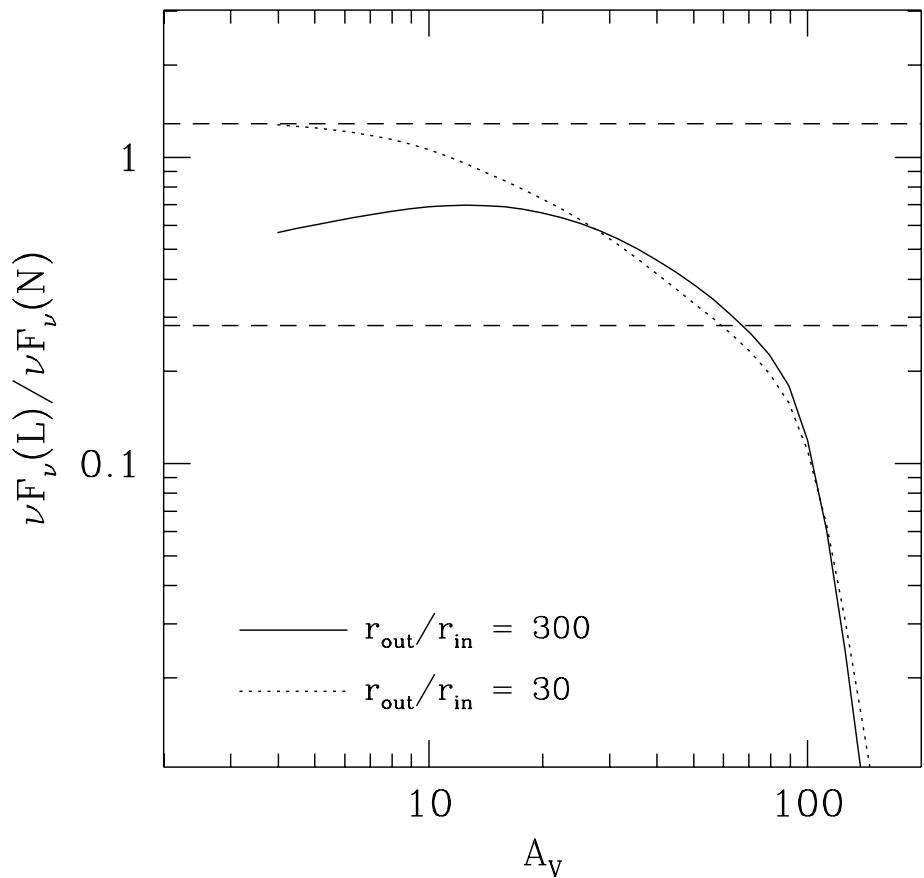}
\vspace*{-0.5cm}
\caption{
The ratio of the L to N band fluxes predicted by the GD model, as a
function of the optical thickness, for two different choices of
$r_{out}/r_{in}$ and an equatorial line of sight.  The two horizontal
lines bound the range observed among nearby Seyfert 2 nuclei. }
\label{figratio}
\end{figure}

Nine out of these galaxies have been observed within a small aperture
both in the near--IR  ($\le 4.5''$)
as well as at 10.6 $\mu$m ($\le 6''$) and clearly detected.  In
Table \ref{tabir} we report the ratio between the L and the N bands 
emission for these narrow--line objects.  In
Fig.~\ref{figratio} the same ratios, predicted by the GD models, are
plotted as a function of the optical thickness $A_V$
and for two values of $r_{out}/r_{in}$. It is clear that the observed IR
spectral slopes imply optical depths along the line of sight 
$A_V\lesssim 80$. Although the sample is not a fair
one and the data are likely to be affected by contributions from
contaminating components, the result suggests that, at least for nearby
objects, the optical depths are broadly distributed but never huge, far
below the value corresponding to Thomson thick matter with a standard
dust to gas ratio. It is apparent from Fig.~\ref{figratio} that 
$A_V\simeq 150$ would correspond to a ratio of the near-IR to mid--IR
emission a factor $\lesssim 0.05$ lower than found in our sample.

A large spread of the optical depths, possibly associated with a
dependence on the luminosity, has been invoked to reconcile the
predictions of unified models to the observed statistics (Lawrence 1991).
As noted by Granato, Danese and Franceschini (1996),  typical IR SEDs of
high-luminosity UVX QSOs are fitted by assuming reprocessing by dusty tori
with equatorial optical thickness 3$\lesssim A_V\lesssim$10, while for
Seyfert 1 galaxies the equatorial optical depths are found in the range
5$\lesssim A_V\lesssim$60. However the equatorial optical depths of 
Type 1 objects are weakly constrained by the available IR and optical 
data. 
\nocite{GDF96}

This might suggest an anticorrelation of the typical optical depths and
of the covering factors with luminosities, although 
significant ranges of thickness and of covering factor
are present at all luminosities.
However Granato et al.\ (1996) showed that the SEDs of hyperluminous IR
high redshift objects and the Cloverleaf, a BAL QSO, can be fitted under
the assumption that the emission is powered by an active nucleus
surrounded by a torus with large equatorial optical depths 60$\lesssim
A_V\lesssim$300. This cleary breaks down the trend for a decreasing
optical depth with increasing luminosity, but may well be linked to a
transient phase when a young nucleus is wrapped in the gas of a young
galaxy (Granato et al.\ 1996).

In summary, only a large range of optical depths can explain the observed
IR spectra of narrow and broad lined AGN. There is no evidence in these
data for circumnuclear tori with extremely large optical depths
($A_V\gtrsim 800$). 

\subsection{Anisotropy of near- and mid-IR emission}
\label{subsecani}

\noindent Heckman (1995), Maiolino et al.\ (1995) and Giuricin et
al.\ (1995) have examined the mid--IR emission of Seyfert 1 and 2
galaxies.  Although the analyses were performed with different methods
a general consensus emerged that narrow--line (Seyfert 2)
nuclei are weaker emitters than broad--line (Sey 1) nuclei. This may be
interpreted as anisotropy resulting from the presence of the obscuring
torus.
\nocite{Hec95,MRR95,GMM95}

In particular Heckman (1995) showed that the average ratio
$S_{10}/S_{1.4}$ of 10.6 $\mu$m flux (within small aperture) to  radio
continuum at 1.4 GHz is about 4 times larger in Seyfert 1 than in type 2
Seyferts. Assuming that the radio emission is indicative of the nuclear
power, he concluded that the putative torus is only mildly anisotropic.
Since the data used by Heckman are rather heterogeneous, the basic
assumption may be not completely safe.  Actually Giuricin, Fadda \& Mezzetti
(1996) found that the total radio fluxes of
Seyfert galaxies correlate with the radio core fluxes, but the total
fluxes are about a factor 6 larger.  Moreover the ratio may be
influenced by  higher level of star formation in Sey 2 hosts (Maiolino
et al.\ 1995), which entails an enhanced radio emission.
\nocite{GFM96}

On the other hand  Heckman (1995) also examined the ratio of 10.6 $\mu$m
emission to the flux in the [OIII] $\lambda$5007 emission line.  The
average 10.6 $\mu$m emission,  normalized to [OIII] flux, of Sey 1s is
larger than that of Sey 2s by a factor of 2.  Although in this case the
result may be affected by absorption and by  possible bias in favour of
small opening angles for Sey 2s, nevertheless we can conclude that the
luminosity at 10$\mu$m is higher by a factor ranging from 2 to 4 in Sey
1s respect to Sey 2s with the same nuclear luminosity.

The samples examined by Heckman (1995), namely the RMS sample (Rush et
al.\ 1993), the CfA sample (Huchra \& Burg 1992), 
and a far-IR selected sample all exhibit the same difference 
between Seyfert 1s and 2s.
\nocite{RMS93,HuB92}

\nocite{MaR95}
For 8 Sey 2s and 9 Sey 1 of the RSA sample, which can be thought as
a volume limited sample, 
10$\mu$m fluxes taken within small aperture as well as radio and [OIII]
fluxes are available. Although this is a small subsample, nevertheless
the analysis confirms that the 10$\mu$m luminosity of the Sey 1s is 
only a factor 2--4 higher than that of Sey 2s with the same nuclear
luminosity.

\begin{figure}
\epsscale{1.0}
\plotone{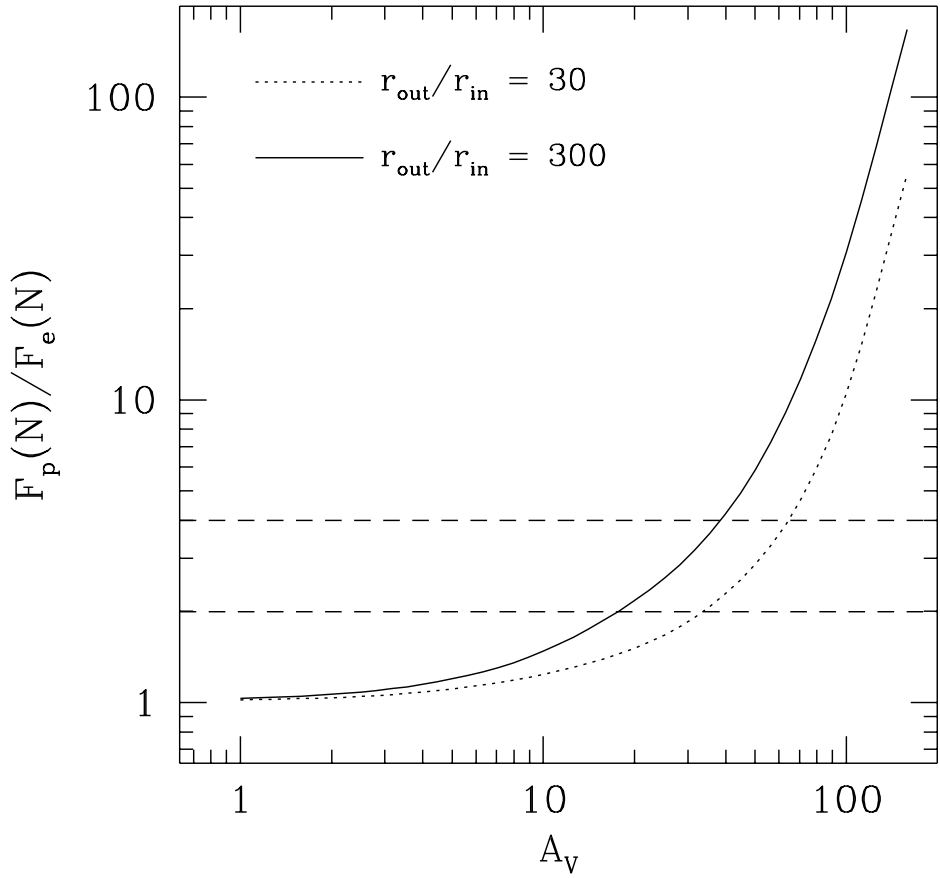}
\vspace*{-0.5cm}
\caption{Anisotropy of the torus mid--IR emission. We plot
the ratio between the fluxes received by a polar and an equatorial
observer at 10.5 $\mu$m, as a function of $A_V$. The two horizontal lines
represent mean values estimated by Heckman (1995) by comparing the
properties of Sy 1 and Sy 2 nuclei.}
\label{figani}
\end{figure}

As it is apparent from Fig.~\ref{figani}, the extended model predicts
that Sey 1s are stronger emitters at 10 $\mu$m than Sey 2s by an average
factor 2-4, provided that the {\it average} absorption along the line of
sight falls in the range $20\lesssim A_V \lesssim 70$, in full agreement
with the fits to available spectral data (see Sect.~\ref{subsecsed}
above). Also figure 4 shows that $A_V\gtrsim 150$ would imply an anisotropy 
8--10 times larger than that observed in the analysed samples.  
The angle $i$ between the torus axis and line of sight toward the
nucleus is expected to be randomly distributed in the RSA sample and in
the CfA sample, though in the latter some bias against very faint nuclei
may be present, affecting the randomness of the inclination.  For a
random distribution half of the nuclei should be seen with an
inclination $i\gtrsim 60^{\circ}$, much larger than the average half
opening angle inferred from the ionization cones $\theta\lesssim
30-40^{\circ}$. Therefore for a significant fraction of Sey
2s of the RSA and CfA samples the line of sight is expected to lie quite
close to the equatorial plane of the torus and, as a consequence, to
exhibit the largest absorption.

The GD torus model predicts that the anisotropy is rapidly
decreasing with increasing wavelengths and, for the typical values of
$A_V$ inferred from the above considerations, practically vanishes at
around 30 $\mu$m (see Fig.~\ref{figexa}). Indeed Mulchaey et al (1994)
\nocite{MKW94} found that the distribution of the ratio of the integrated
IR  flux between 25 and 60 $\mu$m to the [OIII] flux is rather similar
for Seyfert 1 and 2 galaxies, with Sey 2s showing larger variance and a
tail of higher IR to [OIII] ratios, possibly due to starburst components
(see Sect~\ref{subsechsr}).

\nocite{BWG93,CSS93}

In conclusion the mild anisotropy estimated in various samples of
Seyfert nuclei is well reproduced by the extended and moderately thick
tori suggested by the fits to available broad band spectral data
(Sect.~\ref{subsecsed}). By converse, in order to explain the observed
anisotropy, the compact and very thick tori
require a fine tuning of the fraction of inner torus walls directly
seen by the observer, as well as additional emission of extended dust,
possibly associated to the NLR.

\subsection{Infrared spectroscopy of the narrow--line nuclei}
\label{subsecspe}

Infrared spectroscopy can potentially penetrate an obscuring torus and
give direct evidence of the presence of a broad line region completely
obscured at optical wavelengths, where absorption is larger.  Broad
components of Pa$\alpha$ and Pa$\beta$ hydrogen lines have been searched
for in active galaxies.  Actually, the extinction at their near--IR
wavelenghts is a  factor of 3--5 below that at H$\alpha$.
Clear detection has been obtained for several
narrow--line galaxies (Rix et al. 1990; Ruiz, Rieke and
Schmidt 1994; Goodrich, Veilleux and Hill 1994).
The relevant result is that there are optically narrow-line active
nuclei exhibiting broad--lines when observed in the IR domain.
\nocite{RRR90,RRS94,GVH94}

In 3 out 15 Sey 2 nuclei, selected with no particular criterium,
Goodrich et al.\ (1994) were able to detect the broad component of the
Pa$\beta$ line, though their sensitivity limited possible detection to
cases with $A_V\lesssim$11.  In the sample selected by Ruiz et al.\
(1994) on the basis of the relatively strong emission at 3 $\mu$m there
are six out nine nuclei with BLRs.

The detection of a broad component of the Pa$\beta$ in the spectrum of
the 1.9 Seyfert galaxy NGC 2992 suggests $A_V\sim$5--8 (Goodrich et
al.\ 1994; Rix et al.\ 1990).  Similarly in the case of NGC 5506
reddenings $A_V\sim$5--11 from IR broad lines have been reported
(Gooodrich et al.\ 1994; Rix et al.\ 1990).  These reddenings agree
reasonably well with those inferred from the IR colors of the nuclei
(Table~\ref{tabir}).

The existence of tori with modest optical depth is confirmed by
observations of Pa$\alpha$ lines in a complete sample of radio galaxies
(Hill et al. 1996). In three out of eight objects the broad
component of the Pa$\alpha$ has been clearly detected. The values of the
reddening inferred from line ratios are in the range 2.7$\lesssim
A_V\lesssim$7.\nocite{HGD96}

On the other hand larger reddenings of BLRs are surely present in
narrow--line nuclei such as  Cygnus A, exhibiting $A_V\gtrsim$ 24 (Ward
et al.\ 1991), and NGC 1068.
\nocite{WBW91}

These results are the most relevant in discriminating between 
extremely and mildly thick models, 
thanks to their rather different predictions.  At the most basic
level for extremely thick models, such as those
presented by PK, the optical depth at near--IR, where BLRs have been
detected, is so large $\tau_{1 \mu m}\gtrsim $100 to prevent detection
of any IR broad line. In the case of extremely high optical depths
the problem can be solved by assuming that the observed broad-lines 
are scattered in the line of sight by a screen. In this hypothesis 
the lines are expected to be strongly polarized, whilst low level of
polarization is predicted for GD models. Thus IR-spectropolarimetry can
discriminate between the two possibilities.
In addition, in the case of GD model we expect
that narrow-line nuclei with significant emission at near IR are objects
seen along less obscured line of sight and, as a consequence, with a
more easily detectable BLR in the IR. By converse in the PK model there
is no expected correlation between near IR flux (attributed to the
visible portion of the inner walls) and the detectability of the BLR in
the IR domain.

Thus presently available IR spectroscopy data tend to
exclude the general presence of
tori optically very thick even at near IR as a rule, strengthening the
findings of Sects~\ref{subsecsed} and \ref{subsecani}.

\subsection{High spatial resolution observations and
obscuring matter around AGN}
\label{subsechsr}

Valuable information concerning the spatial extent of obscuring matter
and, to a lesser extent, also its column densities, has come in the last
few years from  high spatial resolution observations,
performed with different techniques.

On one hand, water maser line emission observed in the nucleus NGC 4258
at 22 GHz on a subparsec scale (Watson \& Wallin 1994; Miyoshi et al.\
1995; Greenhill et al.\ 1995)  yields clear evidence of the presence of
molecular gas very close to the active nucleus. \nocite{WaW94,MMH95,GHB95} 
On the other hand the images obtained by the Planetary Camera of HST of
the active galaxy NGC 4261 clearly show an unresolved ($<0.1''\simeq 7$
pc) source surrounded by a dusty disc extending on a scale of 100
parsecs (Jaffe et al.\ 1993).  This observation is a rather direct
support to the idea of the existence of extended dusty tori around AGN.
Also imaging in [OIII] and $H_{\alpha}$+[NII] lines with the Hubble
Space Telescope has revealed an unresolved region of very strong
reddening within 23 pc of the nucleus of NGC 2110, a faint Seyfert 2
galaxy (Mulchaey et al.\ 1994).
\nocite{JHF93,MKW94}

$H_2$ maps of the nuclear region of NGC 7469 present evidence of the
presence of molecular gas in a Seyfert 1 galaxy (Genzel et al.\ 1995).
With standard assumption a total gas mass of $\lesssim 10^8$ $M_{\odot}$
is found to be within $\sim 100$ pc from the galaxy centre. This amount
is comparable to that found in a similar volume in NGC 1068 (Genzel at
al.\ 1995). \nocite{GWT95}

High spatial resolution observations of molecular emission in NGC 1068
are available at various wavelengths (Planesas et al.\ 1991, Jackson et
al.\ 1993, Tacconi et al.\ 1994, Blietz et al.\ 1994). These observations
show that the molecular gas extends over 100-200 pc far from the nucleus
with a similar scale height. The associated total extinction may vary from
$A_V \simeq 10$ up to 200. \nocite{PSM91,TGB94,BCD94}

\begin{figure}
\epsscale{1.0}
\plotone{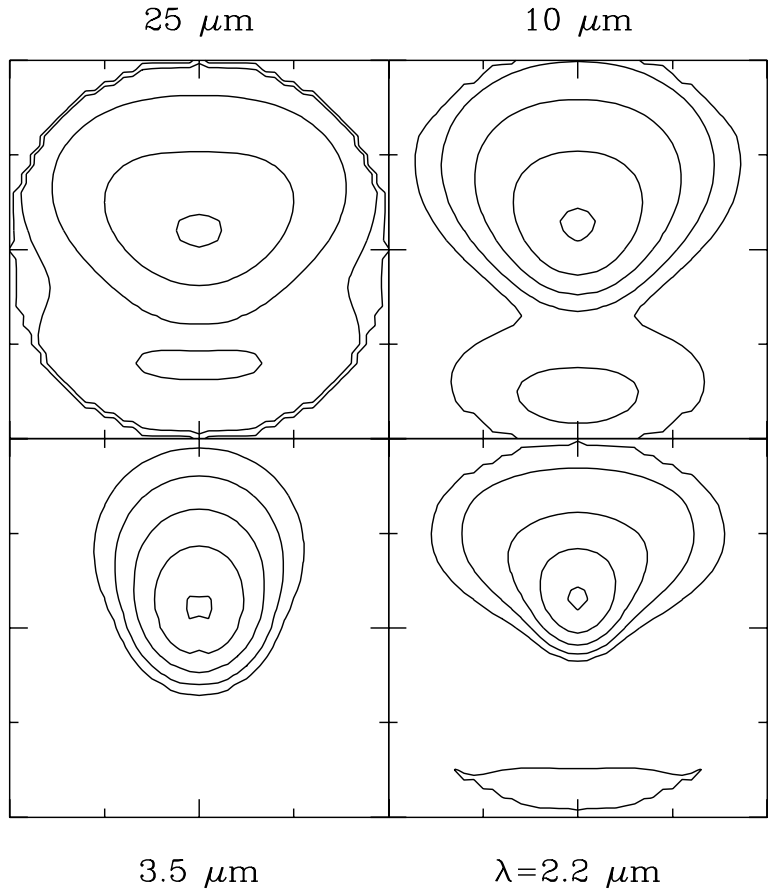}
\vspace*{-0.5cm}
\caption{Brightness contours at four different wavelengths of the
torus used to reproduce the SED of NGC 1068 and observed
from $\theta=65^o$.
The boxes have a width of $2 \, r_{out}\simeq 0.6''$. The maps have been
convolved with a gaussian PSF with FWHM$=0.2 \, r_{out}$ and the levels
refers to 0.01, 0.03, 0.1, 0.3 and 0.9 of the peak.
}
\label{figmap}
\end{figure}

In Figure~\ref{figmap} we present the maps at 2.2, 3.5, 10.5 and 25
$\mu$m of NGC 1068 predicted by the GD model. The size of the torus
emission increases with increasing $\lambda$: the typical radius of the
isophote at 10\% of the peak runs from $\sim 10$ pc in the
near--IR to more than 20 pc in the mid--IR. Bear in mind anyway that the
peak--normalized isophotes depend on the adopted PSF.  The isophotes
are elongated along directions with less absorption, in nice agreement
with observations (Braatz et al.\ 1993; Cameron et al.\ 1993).  However
our model extends over only 0.6$''$, whereas these observations suggest
significant emission up to 1.5--2$''$ at around 10 $\mu$m.
Cameron et al.\ (1993) and Pier and Krolik (1993)
envisaged the possibility that the extended mid--IR emission is due to
dust located in the NLR. However, as pointed out by Cameron et al., this
dust, if diffuse, would  produce large absorption $A_V\sim 3$--4, while
measured values toward the NLR of NGC 1068 are $A_V\lesssim 1.5$
(Inglis et al.\ 1995).
\nocite{BWG93,CSS93,IYH95}

On the other hand, high angular resolution observations also showed that
the star formation rate is significant in circumnuclear regions of a
large majority of AGNs. Evidence of OB star associations in the very
central regions ($r\leq 2$") of NGC 1068 has been found with COSTAR by
Macchetto et al.\ (1994). Also in NCG 7469 robust star formation is present
(Genzel et al. (1995) and in MKN 348 (Simpson et al.\ 1996). 
\nocite{MCS94,SMW96} 
Thus there might be a smooth transition of the mid- and far-IR 
emission from an inner dusty torus
with typical dimensions $r_{out}\sim 50-200 \, L_{46}^{1/2}$ pc,
illuminated by the nucleus and responsible for most of the nuclear
extinction, to a broader dusty region extending over several hundreds of
parsecs, in which the dust is mainly heated by young stars.

The observations summarized in this section therefore confirm that the
dense dusty medium is spread in the nuclear region from a fraction of
pc up to hundreds of pc, yielding absorption $10\lesssim A_V\lesssim
200$.

\section{Relationship of the torus with the X--ray absorption}

\label{secxir}

Seyfert 1 galaxies often exhibit X--ray absorption in excess over the
Galactic column density, although observed hydrogen column densities are
usually $N_H\lesssim 10^{22}\ \rm{cm}^{-2}$ with a median value
$N_H\simeq 10^{21}$ (Nandra and Pounds 1994). This agrees
with the commonplace that in the case of broad--line AGN the line of
sight is relatively free of absorbing material.  By converse hard X--ray
spectra of Seyfert 2s show evidence of larger absorption, as expected in
the unification scheme. \nocite{NaP94}

Spectra detailed enough to allow good estimates of the absorbing column
density are available  for 21 narrow--line active nuclei (see Smith and
Done 1996;  Iwasawa 1995).  For this
sample, admittedly  not a complete one and presumably biased toward low
values of $N_H$, the column density distribution ranges over 21.5$\leq
\log N_H \leq$25 with a median value $\log N_H\simeq$23.3, larger than
that of Sey 1s by a factor 200. \nocite{SmD96,Iwa96}

Although we assume that a fraction of objects with
$N_H\gtrsim 10^{24}$ has been missed, nevertheless these results show
that many of Sey 2s have tori optically thin to electron
scattering (Thomson depth $\tau_T$=1 corresponds to $N_H= 1.5 \times
10^{24}$ and $A_V$=750). 

Can the X--ray absorbing hydrogen column be associated to the dusty
torus responsible for the UV and optical absorption and for the IR
emission? Assuming standard dust over hydrogen abundance, the median
$\log\ N_H\simeq$23.3 translates into a median reddening $A_V\simeq$100.
This would imply that the Seyfert 2s on the average should be fainter at
10$\mu$m than Sey 1s by a factor of $\gtrsim 15$ (see figure 4), 
much larger than the
factor 2--4 found by Heckman (1995). Indeed in Sect.~\ref{subsecani} we
have shown that this lower anisotropy of the torus emission, confirmed
by our analysis of the RSA sample, requires $10\lesssim A_V \lesssim
70$.  Host galaxy can not help much in accounting for this discrepancy,
since the objects have been observed with small aperture (a few
arcseconds).  Defining $A_V^{X}$ as the absorption in the V band that
associated to the X--ray absorbing column density $N_H$ under the
assumption of normal gas--to--dust ratio, and $A_V^{IR}$ the
absorption derived from the IR spectra, we may infer that on the average
$A_{V}^{IR}\sim 0.1-0.5 \, A_{V}^{X}$.

The discrepancy is confirmed by the analysis of a sample of objects
with accurately determined X--ray spectra, O[III] line fluxes and
10 $\mu$m fluxes. The relevant data are reported in Table~\ref{tabx}.
The twelve Sey 1s of the sample show a rather narrow distribution of the
ratios of the X--ray to [OIII] flux, $40\leq F_{2-10}/F_{5007}\leq 360$
(fluxes expressed in erg  s$^{-1}$ cm$^{-2}$), smaller values being
associated with higher column densities.  For the same objects the
spread of the distribution of $F_{10\mu m}/F_{5007}$ is limited to a
factor of 9, which is smaller than the factor of $\sim$ 40 found by
Heckman (1995).  The effect is attributable to the fact that the
contamination by underlying galaxy is lower in our sample, since the
objects are relatively nearby.  These results confirm that the
normalization to the intrinsic nuclear luminosity through the [OIII]
emission is appropriate, though absorption and use of different
apertures in the observations clearly introduce noise.

The median value for the 12 Seyfert 1s is $F_{2-10}/F_{5007}\simeq 60$,
larger than that of the 11 Seyfert 2s by a factor 2. With a numerical
code which solves the transport equation for X--rays, taking into
account both photoelectric absorbtion as well as Compton scattering
(Granato 1997, in preparation), we found that a decrease of a factor 2
implies an
average column density $N_{H}^{X}\simeq 1\times 10^{23}$ for the
obscured objects, only a factor 
2 smaller than the median column density found in the Sy 2s sample. 

If the material responsible for the X--ray absorption had a normal
gas--to--dust ratio, we would expect $A_{V}^{X}\simeq 50$ and a ratio of
IR to [OIII] fluxes which is at least a factor of 4 smaller for
narrow--line active galaxies, while the ratio of the IR to
[OIII] fluxes is rather similar for the broad and narrow--line galaxies
of the sample. 

\begin{figure}
\epsscale{1.0}
\plotone{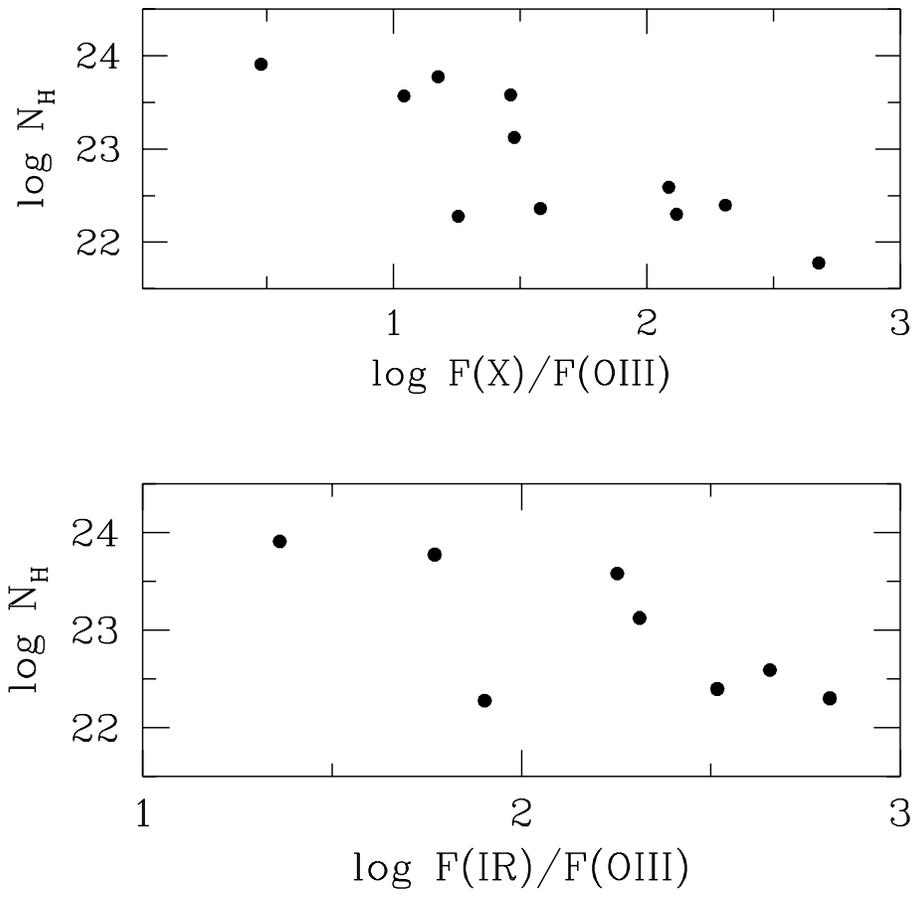}
\caption{Correlation between the $N_H$ derived from X--ray spectral fits
and the X and IR luminosities normalized to the OIII luminosity.}
\label{figxnh}
\end{figure}

Good correlation exists between the ratio of 2--10 keV
flux $F_{2-10}$ to the O[III] line flux $F_{5007}$ and the X--ray
absorbing column density of the Sy 2s sample (see Fig.~\ref{figxnh}).
Similar level of correlation is also present between the column density
and the ratio of the 10$\mu$m fluxes $F_{10 \,\mu m}$ to $F_{5007}$
for the 8 objects for which IR measurements are available.
This result is 
understood if we assume some degree of correlation between the X--ray 
absorbing column density and the optical depth due to the dust.

\begin{figure}
\epsscale{1.0}
\plotone{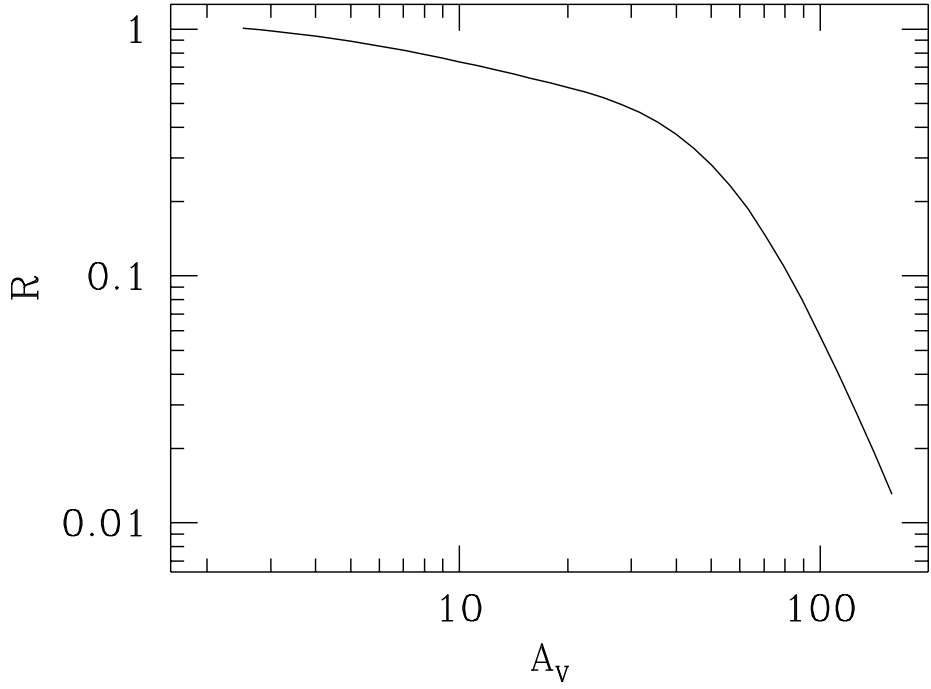}
\vspace*{-1.5cm}
\caption{
$R$ is the ratio $[f_X/f_{12}]_{\rm{Sy 1}} / [f_X/f_{12}]_{\rm{Sy 2}}$
predicted by our radiative transfer codes for IR and X--ray, when
$A_{V}^{IR}$ is forced to be equal to $A_{V}^{X}$.
Observationally $R>1$, while for $A_{V}^{X}  \simeq 100$,
the mean observed value, $R$ should be $\sim 0.05$.
To solve the discrepancy $A_{V}^{IR} \sim 0.2 \, A_{V}^{X}$ is required.
}
\label{figrapxir}
\end{figure}

The 15 Seyfert 1 galaxies of the sample in Table~\ref{tabx} exhibit
ratios $0.16 \le F_{2-10}/F_{10 \mu\mbox{m}} \le 0.7$ with a median
value 0.5, while for the 8 Seyfert 2s the ratios are significantly clustered 
around 0.2. Hence the Seyfert 1 galaxies have an X-ray to mid-IR average 
ratio larger by a factor of about 2 than that of Seyfert 2s.
On the other hand in any reasonable model in which $A_{V}^{IR}$ is 
forced to be equal to $A_{V}^{X}$ the ratio
$$
R=
\frac
{ [F_{2-10}/F_{10\mu m}]_{\rm{Sy 1}} }
{ [F_{2-10}/F_{10\mu m}]_{\rm{Sy 2}} }
=
\frac
{ F_{10\mu m,\rm{Sy 2}} / F_{10\mu m,\rm{Sy 1}} }
{ F_{2-10,\rm{Sy 2}} / F_{2-10,\rm{Sy 1}} }
$$
turns out to be well below 1 (Fig.~\ref{figrapxir}).
In particular with $A_{V}^{X} = A_{V}^{IR} \simeq 100$ (the median
 $A_{V}^{X}$),
our radiative transfer codes predict $R \simeq 0.05$. 
Once more, the discrepancy is solved if $A_{V}^{IR} \sim 0.1 \, A_{V}^{X}$

The link of the X--ray to the mid--IR emission in AGNs has been
investigated also by Barcons et al.\ 1995), who measured the X--ray
intensities in the nominal band 2--10 keV, as observed by the experiment
A2 on board of HEAO--1 satellite, at the positions of the 
RMS sample of AGNs selected at 12 $\mu$m by IRAS.
After subtracting the "blank sky" they
found that the ratio of the flux at 5 keV to the flux at 12 $\mu$m
$f_X/f_{12}= 1.4_{-0.4}^{+1.1} \times 10^{-6}$ for the 54 Seyfert 1s and
$f_X/f_{12}= 2.0_{-1.5}^{+4.5} \times 10^{-7}$ for 59 Seyfert 2 galaxies
a factor of 7 lower. The low angular resolution
of the IRAS data implies that the host galaxies contributes to the 12
$\mu$m fluxes, in a way depending on the relative strength of the galaxy
to the nucleus as well as on distance. Maiolino et al.\ (1995) claim
that the typical host galaxy of a Seyfert 2 nucleus is 5 times more
luminous that the typical Sey 1 host galaxy.  Thus part of the factor of
7  found by Barcons et al.\ (1995) as difference in the X--ray to IR
ratio may be accounted for by this effect. However also for this 
IR selected sample the ratio of the X-ray to the mid-IR emission
\ is larger for type 1 Seyferts, while, assuming $A_{V}^{X}=A_{V}^{IR}$, 
the opposite is expected. \nocite{BFD95}

It is worth noticing that the median value $F_{2-10}/F_{10 \mu m}
\simeq 0.5$ found for the Seyfert 1 galaxies of Table~\ref{tabx} translates
into $f_X/f_{12}\simeq 6 \times 10^{-6}$, a factor about 4 larger than
the value given by Barcons et al.\ (1995). On the other hand the
sample of Table~\ref{tabx} can be thought as X-ray selected and 
thus is biased toward larger X-ray to IR ratios. If the parent
population has a dispersion $\sigma\sim 0.4$ in the distribution of the
logarithms of the
infrared to X-ray luminosity ratios, the factor 4 is easily recovered
(see eq. 14 of Barcons et al.\ 1995).

For an handful of objects the absorption in the visual band $A_{V}^{X}$,
calculated under the assumption that the X--ray absorbing gas has a
normal gas--to--dust ratio, can be directly compared to that derived
from the IR spectrum $A_{V}^{IR}$ and/or to the absorption derived from
IR broad lines $A_{V}^{BL}$.  NGC 2992 has low $A_{V}^{X}\simeq$11 and
$A_{V}^{IR}$ is not well determined but possibly small, $\simeq$ 5--11.
It is worth mentioning that Weaver et al.\ (1996) proposed \nocite{WNY96}
that the Fe $K_{\alpha}$ fluorescence line in this source is reprocessed
at distance of 10$\pm$4 l.y. from the central source in a region with
$N_H\simeq$2--4$\times 10^{23}$. The argument is based on long term
variability of the line equivalent width of a factor 3, compared to a
continuum variation of a factor 20. However the poor time statistic does
not exclude a significantly smaller distance of the reprocessing
material. Similarly in NGC 5506 $A_{V}^{X}\simeq$19, the IR broad band
spectrum demands low absorption, and $A_{V}^{BL}\simeq$5--11. 
A significant change in the relationship between X--ray and IR
absorptions is apparent at higher X--ray column density. Indeed, NGC
4388 $A_{V}^{X}\simeq$ 190, while $A_{V}^{IR}\simeq 15$ and IR
broad--lines have not ben detected. The IR spectrum of MKN 348 suggests
$A_{V}^{IR}\simeq$7--10 (Roche et al.\ 1991) and indeed the
broad component of He I (1.083 $\mu$m) line has been detected (Ruiz et
al.\ 1994). \nocite{RAS91,RRS94}

For the same object Smith and Done (1996) estimated $N_H\simeq
1.3 \times 10^{23}$, corresponding to $A_{V}^{X}\simeq$70. In NGC 1068
the GD model of the IR spectrum requires $A_V^{IR}\simeq 72$ along the
line of sight, while X-ray spectral observations suggest $A_V^{X}\geq
4000$, almost 500 times larger.  Similar conclusions holds for
Circinus: the beautiful spectrum obtained with the SWS on board of ISO
by Moorwood et al.\ (1996) demands $A_V^{IR}\sim 50$, while the X--ray
spectrum obtained by Matt et al.\ (1996) with ASCA implies $N_H\gtrsim
10^{24}$ at least and thus $A_V^{X}\gtrsim 500$.

Therefore for the objects of Table~\ref{tabx} for which reliable
determinations of X--ray and IR absorption are available, we find
$A_{V}^{IR}\sim 0.002-1 \, A_{V}^{X}$, the equality holding only for
objects with low $N_H$.

This behaviour is confirmed by high luminosity objects, which can
be identified with type 2 QSOs.  The IR and optical spectrum of IRAS
09104 is nicely fitted by a GD torus with an absorption along the
line of sight $A_{V}^{IR}\simeq 11$ (Granato et al.\ 1996), whilst the
X--ray spectrum indicate that only scattered radiation is seen with ASCA
(Fabian et al.\ 1995), implying $N_H>>1.5 \times 10^{24}$ and, as a
consequence, $A_{V}^{X}>>$800.  For IRAS 10214 the line of sight
absorption is $A_{V}^{IR}\simeq 150$ while X-ray upper limits suggest
$A_{V}^{X}\gtrsim 1500$ (Granato et al.\ 1996). \nocite{FaC95,GDF96}

In conclusion available data, combined with GD model, implies that
$A_{V}^{IR}\sim A_{V}^{X}$ at low X--ray column density $N_H\lesssim
\rm{few} \times 10^{22}$, while when $3\times 10^{22} \lesssim N_H \lesssim
10^{24}$ we have $A_{V}^{IR}\sim  0.1 \, A_{V}^{X}$. Finally
$A_{V}^{IR} << 0.1 \, A_{V}^{X}$ for the largest measured column
densities $N_H\gtrsim 10^{24}$

\subsection{Interpretations}

Although the overall picture may be quite complex and data may be
plagued by systematic effects such as
X--ray variability, nevertheless there is evidence that the
gas responsible for the X--ray absorption is in general dust free and
that the dusty torus in Seyfert 2 galaxies is not the site wherein the
bulk of the X--ray absorption occurs. Only in the least absorbed objects
we find that $A_V^{IR}\sim A_{V}^{X}$. Moreover in the data there is no
significant trend of the ratio of the X-ray to IR absorption in
type 2 AGN on the luminosity. The increasing number of claimed
detections of type 2 QSOs (see e.g.\ Boyle et al 1995;
Almaini et al 1995; Ohta et al.\ 1996) is
suggestive of a weak, if not null, correlation of the absorptions
themselves with luminosity. \nocite{BMW95,ABG95,OYN96}

The most natural explanation of these results is that the X-ray
absorption occurs in gas which lies inside the dust sublimation radius.
The nuclear activity implies a reservoir of ISM, which flowing toward
the nucleus fuels the central BH.  We identify the reservoir with the
dusty extended torus and we relate the X--ray absorbing structure with
the material which, on the way from the reservoir to the accretion disc,
is already beyond the dust sublimation radius.  Assuming that inside
this radius the gravity is dominated by the central BH and that a quasi
steady--state is established, we can derive the expected density by mass
conservation
$$
\rho={\dot M\over 4\pi \, r^2 v_r}
$$
\noindent where $\dot M$ is the accretion rate and $v_r$ is the radial
drift velocity.  If we introduce the ratio of the drift to the circular
velocity $k=v_r/v_{\phi}\sim 0.01-0.1$ (see e.g. Blackman and Yi 1996; 
Krolik and Begelman 1988) and we assume
that the circular velocity is nearly keplerian, we get for the hydrogen
number density \nocite{BlY96,KrB88}
$$
n_H \simeq {1\over 4 \pi \, k \, m_H G^{1/2}} {\dot  M\over r^{3/2} 
M_{BH}^{1/2}}
$$
\noindent The sublimation radius is $r_{in}\simeq 0.5 \, L_{46}^{1/2}$
pc, about a factor 2--3 larger than the BLR typical radius. In the
intermediate region dust is disrupted, but the metal rich gas absorbs the
X--ray flux coming from the nucleus. Let this region extend from
$f \times r_{in}$ to $r_{in}$ and  $F=L/L_{Edd}$.
Then the expected column density of this region is
$$
N_H\simeq {0.8\times 10^{23} \over k} (f^{-1/2}-1) \,
\epsilon_{0.1}^{-1} F^{1/2} L_{46}^{1/4} \ \ \ \mbox{cm}^{-2}
$$
\noindent where $\epsilon_{0.1}$ is the efficiency of mass into
radiation conversion normalized to 0.1 and the radial integration
assumes a constant $k$. Inserting the reasonable values $f\simeq 0.5$
and $\epsilon_{0.1}\simeq 1$ we get for Seyfert nuclei with luminosity in the
range $10^{44}-10^{45}$ and emitting near the Eddington limit
$10^{23}\lesssim N_H\lesssim 10^{24}$ for $k\simeq 0.01-0.1$.
Although this view is admittedly crude, we expect that at about the
sublimation radius radiation pressure, differential
rotation, viscosity and gravitational field are arranged in a way to
allow gas feeding the central black hole (see e.g.\ Krolik and Begelman 1988;
Yi et al.\ 1994; Blackman and Yi 1996;
Granato et al.\ in preparation). \nocite{BlY96,KrB88,YFB94,BlY96}

Interestingly enough the dependence on the total luminosity is
relatively weak, while it is more pronounced that on $F=L/L_{Edd}$.
The column densities observed in Seyfert 2 galaxies
suggest that they are emitting at the Eddington limit.

In conclusion from the analysis of available IR and X--ray data, it is
apparent that the X--ray absorption in type 2 AGN is chiefly due to
matter free from dust, located at and just inside the dust sublimation
radius and flowing toward the accretion disc.

\section{AGN statistics and the X--ray and IR background}
\label{secstat}
\noindent

The relative fraction of broad--lined and narrow--lined AGN is extremely
relevant to a number of problems such as the validity of the unified
schemes (see Lawrence 1991), the origin of the X--ray background (see
e.g. Setti and Voltjer 1989) and the contribution of the AGN
to the IR background (Granato et al.\ 1995). Unfortunately
the available statistics mainly refer to the local populations, while
for the X--ray and IR backgrounds the relevant statistics concern objects
at substantial redshift $z\gtrsim$1. On the other hand we may derive
interesting clues also from local samples. \nocite{SeW89,GFD95,Law91}

Lawrence (1991) \nocite{Law91} 
showed that the fraction of local narrow--line AGN  may
range from  0.7 to 0.84 in optically selected samples, after corrections
for bias.  Of course the statistics is rather poor for faint objects
even in the case of local samples. To overcome this bias Maiolino and
Rieke (1995) \nocite{MaR95} 
examined the RSA sample and found that the fraction of type 2
nuclei may be as large as 0.8, with an important fraction of low luminosity
objects.

Assuming that the relative number of AGN merely reflects the average
opening angle $\Theta_H$ of the dusty torus, the optical
samples suggest $35^o\lesssim \Theta_H\lesssim 45^o$, a range
confirmed also by the observations of ionization cones.

The large RMS sample of 116 Seyferts selected at 12 $\mu$m exhibits an
almost identical fraction of type 1 and type 2 objects. 
To understand this result we have to take into account
that on the average type 1 objects are brighter by a factor of 2--4
(Heckman 1995) and that  the host galaxy
contribution is not negligible for the objects in the RMS sample,
because of the poor angular resolution of IRAS.
Indeed under the assumptions that locally 
the host galaxy is as luminous as
the active nucleus in type 1 objects at 12 $\mu$m and three times 
more luminous in type 2 objects, and that type 2 objects 
are four times 
more numerous than type 1, we predict almost the same number of the 
two types at the IRAS 12 $\mu$m survey flux limit.

The statistics on the X--ray absorption is still rather poor. The $N_H$
distribution of the Seyfert 2s {\it detected} in hard X-ray bands (Smith
\& Done 1996) shows that a significant fraction of about 40$\%$ of the
objects has $N_H<10^{23}$ and only 15$\%$  has $N_H>10^{24}$. This
result must be taken with caution, as it is surely plagued by bias
against highly absorbed objects. However, following the conclusions of
the previous section, surveys in the mid--IR bands will
not miss many of the higly X--ray absorbed AGN and will be extremely
helpful in settling down the relative fraction of type 1 and 2 objects.

This fact is extremely relevant to the problem of the X--ray background.
As proposed by Setti and Voltjer (1989) the hard X--ray background from
3 to 100 keV (HXRB) can be fitted by the integrated emission of heavily
absorbed AGN (see also Zdziarski, Zycki and Krolik 1993;
Madau, Ghisellini and Fabian 1993; Comastri et al 1995;
Celotti et al 1995).  \nocite{SeW89,ZZK93,MGF93,CSZ95,CFG95}
Although with some differences,
all these authors concluded that the HXRB can be produced by an evolving
population of obscured AGN, characterized by local luminosity function
and local volume emissivity within the observational boundaries, and
endowed with a significant cosmological evolution up to
$z_{max}\sim$2.5-5 ($\langle nL \rangle \propto (1+z)^C$ with
$C=2.2-2.7$).  The obscured AGN are required to be 2--4 times more
numerous than the unobscured ones. About 50\% of the obscured AGN are
required to have $N_H>5\times 10^{23}$ and about 35\% to have $1\times
10^{24}\leq N_H \leq 5\times 10^{24}$ (see e.g.\  Celotti et al.\ 1995).
These requirements are consistent with the limits imposed by
observations, although not by a wide margin as for the column density
distribution (see Smith and Done 1996).

A viable alternative solution for the HXRB has been worked out by
Franceschini et al.\ (1993), \nocite{FMD93} 
who proposed, on the basis of the available
X--ray source counts and related statistics, that the HXRB is mainly
contributed by a population of AGN endowed with luminosity and spectral
evolution (luminosity and column density increasing with increasing
redshift).  Although both schemes imply that a significant number of
absorbed AGN contributing to the HXRB will be detected with the ISO
surveys, nevertheless the different assumption about spectral evolution
will be mirrored in the IR surveys with different redshift distribution
of the type 2 AGN.

\begin{figure}
\epsscale{1.0}
\plotone{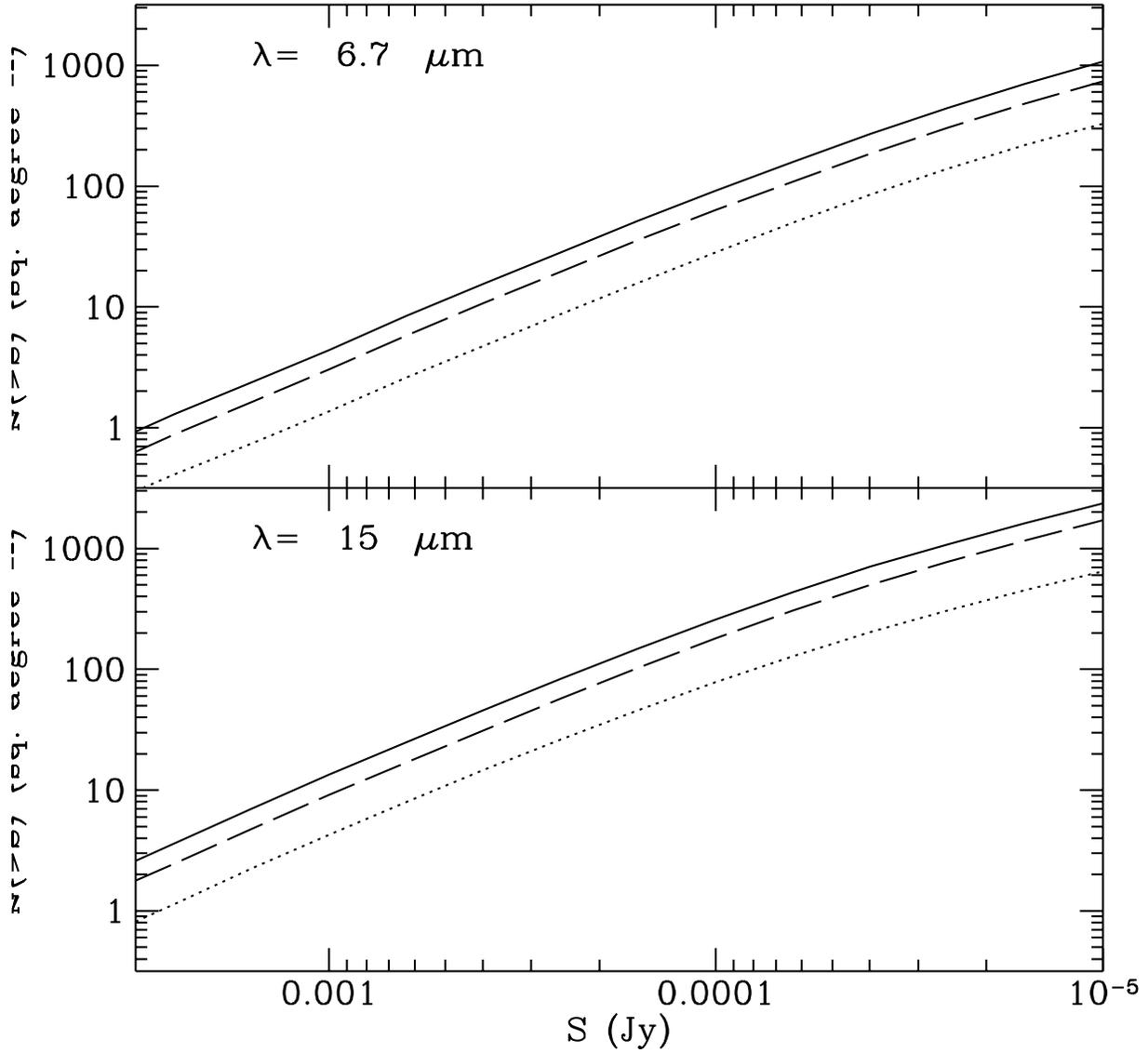}
\vspace*{0.5cm}
\caption{AGN counts predicted in mid-IR bands under the requirement 
that they produce the hard X-ray background. The dashed line 
and the dotted line refer to obscured and unobscured nuclei
respectively.
}
\label{figcounts}
\end{figure}

In Fig.~\ref{figcounts} and in Table~\ref{tabcounts} 
we present the AGN counts at 15 $\mu$m and 6.7 $\mu$m, the
two effective wavelengths of the ISOCAM surveys, computed with the
prescription that they produce the HXRB. In order to work out IR count
predictions the ratio of the 2--10 keV luminosity $L_{2-10}$ to the
luminosity in the 10 $\mu$m band $L_{10\mu m}$ of the unabsorbed objects
must be specified. The 12 type 1 objects reported in Table~\ref{tabx} show an
average value of 0.5 of this ratio.
A similar value $\simeq $0.4 may also be inferred by SED of the UVSX QSO
sample reported by Elvis et al.\ (1994). \nocite{EWM94} We used the results of
Sect~\ref{secxir} to infer the dust absorption in the torus at given
hydrogen column density. In the total flux a contribution from a non
evolving host galaxy is also included as described above.

From Table~\ref{tabcounts}
it is apparent that the ISOCAM surveys will be extremely efficient in
detecting the dusty AGN. They will provide very sound statistics
allowing to understand both the relative number of type 1 and 2 in the
local Universe as well as their respective evolution.

\begin{figure}
\epsscale{1.0}
\plotone{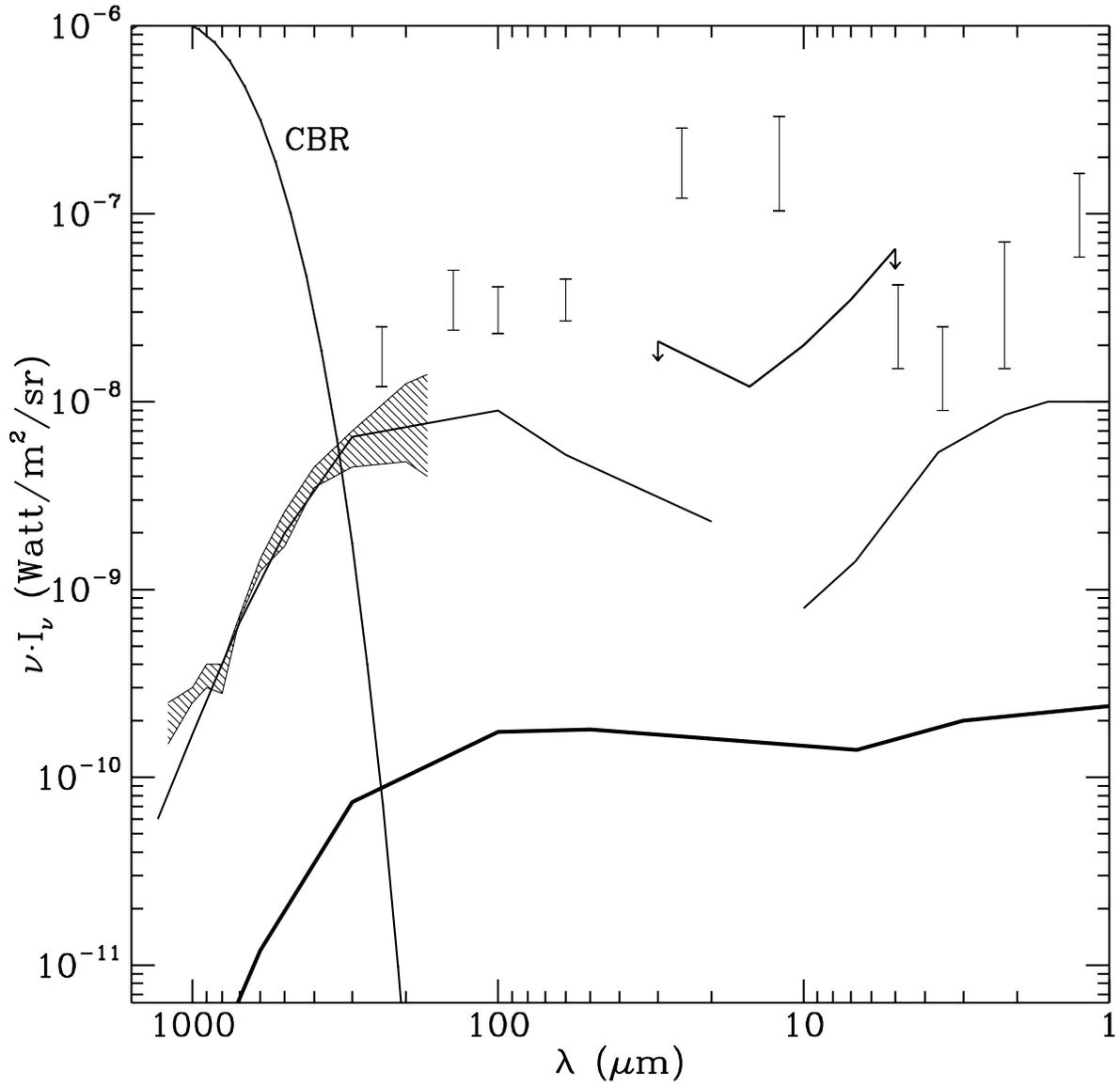}
\vspace*{0.5cm}
\caption{
Contribution of AGNs to the IR background (thick line), compared with
the predicted emission of galaxies (Franceschini et al.\ 1994), 
the tentative sub-mm detection by Puget et al.\ (1995), shaded region)
and available limits by Hauser (1994).
}
\label{figcibr}
\end{figure}

Using the IR counts we can compute the contribution to the IR background
of the objects producing the HXRB, which turns out to be rather flat
from several microns to 200 $\mu$m at level of $\simeq 2\times 10^{-10}$
Watt/$m^{-2}$/sr (see Fig.~\ref{figcibr}). This is only a lower limit to
the AGN contribution, since it is based only on hard X--ray emitting
objects. It is worth noticing that this corresponds to about 5 times the
local energy density of the XRB $\simeq 8\times 10^{-17}\ \mbox{erg}
\, \mbox{cm}^{-3}$ from 1 keV to MeV and still is
only a fraction of the IR background produced by normal galaxies (see
e.g. Franceschini et al.\ 1995). \nocite{FGM95}

The request of modelling the HXRB spectrum summing up absorbed sources
implies that only a small fraction of the total energy is stored in the
HXRB. To illustrate the problem we can compute the BH mass density
required to supply the XRB. This is rather easily done for the model
proposed by Celotti et al.\ (1995), who proposed that absorbed AGN share
the same 2--10 keV luminosity function and the same luminosity evolution
of the unabsorbed ones, but are a factor around 2.5 more numerous.
Using the 47 SEDs of the UVSX sample of QSOs (Elvis et al.\ 1994) the
average bolometric correction of the 2-10 keV band is $k_{bol} \simeq
38$ with very small dispersion. Since the sample includes 29 radio quiet
and 18 radio loud objects the value found possibly underestimates the
bolometric correction for radio quiet objects. Finally the mass density
of the black holes associated to the X-ray background is
$$\rho_{BH}\simeq 4\times 10^{5} \epsilon_{0.1}^{-1}
\frac{k_{bol}}{38} \,\, M_{\odot}\,
Mpc^{-3}$$
\noindent where $\epsilon_{0.1}$ is the mass to energy conversion
efficiency normalized to 0.1. This value has to be compared to $ 
\rho_{BH}\simeq 1.4\times 10^{5} \epsilon_{0.1}^{-1}\ M_{\odot}\ Mpc^{-3}$
found by Chokshi and Turner (1992) for the optical selected AGN. 
By using the galaxy local luminosity function  by Efstathiou et 
al.\ (1988) \nocite{ChT92,EEP88} we find that the production of the XRB 
implies that almost all the galaxies with $M_B<$-17 should
harbor a BH with $M\geq 4\times 10^{7}\ M_{\circ}$. The
consequences and implications will be examined elsewhere.

\section{Summary and conclusions}
\label{secsum}

The statistics of large samples of AGN (CfA, RSA, RMS and others)
on the differences of 10 $\mu$m nuclear emission between broad and
narrow line AGN show that the anisotropy amounts to a factor of 2-4,
implying that the narrow line objects of these samples have on
average dust absorption in the range $10\lesssim A_V \lesssim 80$.  The
detection of broad components of IR permitted lines such as
Pa${\alpha}$ in a number of narrow line active nuclei confirms that low
optical depths are not an exception.  Available data on IR spectra of
AGN strongly support the same conclusion, adding evidence for a
significant spread of $A_V$ in narrow line objects.

Since we would expect in the CfA and RSA samples also objects seen along
lines of sight close to the equatorial plane, these results imply that
the absorption due to dust in the torus is actually significantly
smaller than predicted by PK torus models with $\tau_T \simeq 1$
and $A_V \sim 800$. GD extended torus models are
naturally in agreement with these results. For instance the fit to NGC
1068 nuclear IR spectrum is obtained with $A_V\simeq 72$ along
the line of sight. Moreover the predicted maps of the mid--IR emission
are extended and thus in good agreement with the observation, though
additional emission from the regions just outside the torus seems to be
requested.  This additional emission has been already observed in
several cases, and attributed to the star formation activity, which is
quite natural in these astrophysical settings.

Interestingly enough also the IR broad band spectra of broad line AGN
can be fitted by models with 
dust absorption in the equatorial plane confined at $5\lesssim
A_V \lesssim 50$. The largest values of dust absorption $A_V
\simeq 200$ are required when fitting high redshift and high luminosity
narrow line objects (but also broad line ones such as the Cloverleaf).
These active nuclei plausibly are in a phase of high activity
and have a large reservoir of mass available to the accretion, which
is also responsible for a large covering of the nuclear region.
These cases look rather extreme and notwithstanding they exhibit
dust absorption significantly less than that implied by models 
with $A_V \gtrsim 800$.

The emerging picture suggests that the strictly unified scheme does not
hold. There is evidence for large spread in optical depth 
distribution in both type 1 and type 2 nuclei. Also the
average covering factor may change from obscured to unobscured AGN.
Dependence of these parameters on luminosity is not evident, while
there are cases of objects with quite different luminosities exhibiting
high absorption and high covering factors. We suggest that these 
parameters depend more on the conditions of the interstellar medium
of the host galaxy than on the nuclear luminosity. 

Since the IR observations can be fitted by 
GD torus models predicting a relatively low $A_V$ for the dust, the
next question was about the origin of the X--ray absorption in narrow
line AGN.  The data show a rather strict relationship among the column
density of the X--ray absorbing material and the dust absorption in the
torus, for several cases in which we can observationally infer both of
them.  However we found that $A_V^{IR}\sim A_V^{X}$ only for low
hydrogen column densities $N_H \lesssim \mbox{few} \times 10^{22}$,
while for $3\times 10^{22} \lesssim N_H \lesssim 10^{24}$
$A_V^{IR}\sim 0.1 A_V^{X}$ and for $N_H\gtrsim 10^{24}$ $A_V^{IR}<< 0.1 
A_V^{IR}$. 

We propose that the X--ray absorption takes place mainly in the
region just inside the dust sublimation radius, where dust is disrupted
but a metal enriched gas exists. In this scheme the dusty torus reflects
more the conditions of the reservoir of the material avilable for the
accretion, while the X--ray absorbing column density reflects the
conditions in a region which is dominated by the BH physics. Indeed we
showed  that the X--ray column density can be related to the ratio
$L/L_{Edd}$ at which the nucleus is radiating, higher $L/L_{Edd}$
implying higher column density.

The conclusion that the dusty tori are not so thick to absorb the mid--IR 
nuclear radiation implies that mid--IR surveys should be able to detect
a large number of narrow line AGN, particularly those with $N_H\lesssim
2\times 10^{24}$, which are supposed to be the main contributors to the
HXRB. In particular we have shown that with the results of the ISO
surveys, which are soon being completed, it will be possible to assess
the problems of the local fraction and cosmological evolution of narrow
line AGN. Also the problem of the contribution of the AGN to the IR
background has been examined. We found that even in the case of a large
number of absorbed sources as implied by the HXRB spectrum, the CIBR is
still a fraction of that expected by normal galaxies. Nevertheless we
predict significant correlation of the HXRB and the CIBR particularly at
around 10--20 $\mu$m, where the expected ratio between the background 
produced by the AGN and that due to galaxies is at a maximum.
Also the assumption that the HXRB is produced by absorbed AGN implies
a large mass density in residual black holes 
$ 
\rho_{BH}\simeq 4\times 10^{5} \epsilon_{0.1}^{-1}\ M_{\odot}\ 
\mbox{Mpc}^{-3}
$

\acknowledgments
This work was supported in part by MURST and ASI. We thank the referee, Piero
Madau, for carefully reading the manuscript and for valuable comments.



\newpage

\nocite{PBB96,FMD94}
\newpage

\begin{table}
\begin{tabular}{lccccc}
Name    &  D  & F(L)&  F(N) &  Ref  &  Ra  \\
	  & Mpc  & mJy &  mJy  &       &      \\
Circinus    &   3 &   701 &   6000 &   R,M    &    0.35 \\       
MCG-5-23-16 &  48 &    79 &    530 &   R,R    &    0.44 \\       
NGC 1068    &  22 &  1700 &  17800 &   RL,RL  &    0.28 \\       
NGC 1386    &  25 &    33 &    209 &   A,G    &    0.47 \\       
NGC 2110    &  45 &    47 &    198 &   R,R    &    0.71 \\       
NGC 2992    &  45 &    74 &    259 &   Mi,G   &    0.85 \\       
NGC 4388    &  26 &    51 &    305 &   A,R    &    0.50 \\       
NGC 5506    &  42 &   313 &    720 &   R,R    &    1.29 \\       
NGC 7582    &  32 &   201 &    877 &   K,R    &    0.68 \\       
\end{tabular}
\caption{Near and Mid--IR fluxes of nearby Seyfert 2 galaxies within
small apertures. The photometric data are taken from various papers as
follows: A: Alonso-Herrero et al.\ 1996; 
K: Kotilainen et al.\ 1992; G: Giuricin et al.\ 1995; M:
Maiolino et al. 1995; Mi: Mizutani et al.\ 1994; R: Roche et al.\ 1991;
RL: Rieke \& Low 1975. The first code refers to L bands fluxes, the
second one to N band flux. The last column is the ratio between $\nu
F_\nu$ in the L band and N band.
}
\label{tabir}
\end{table}
\nocite{GMM95,MRR95,RAS91,AWK96,KWB92,MSM94,RiL75}

\newpage
\begin{table*}

\begin{tabular}{lccccccccc}
Name       &$N_H$&$F_{2-10}$&$F_{5007}$&$F_{10 \mu m}$&
Ref X&Ref IR&X/IR& X/OIII& IR/OIII\\
\multicolumn{10}{c}{Seyfert 1}\\
\hline
IC 4329    & &  12.3 &  34 &  22.0 & NP & G  & 0.56 &  362 &   647 \\
MGC-6-30-15& &   5.6 &     &   8.2 & NP & G  & 0.68 &      &       \\
Mrk 1040   & &   2.6 &  13 &   9.4 & M  & G  & 0.28 &  200 &   723 \\
Mrk 335    & &   1.3 &  23 &   6.0 & NP & G  & 0.21 &   54 &   260 \\
Mrk 509    & &   5.0 &  81 &   6.8 & NP & M  & 0.74 &   62 &    84 \\
Mrk 79     & &   2.5 &  37 &   7.2 & M  & G  & 0.35 &   68 &   195 \\
NGC 1365   & &   2.6 &     &  10.0 & M  & G  & 0.26 &      &       \\
NGC 3227   & &   4.0 &  64 &   7.9 & NP & G  & 0.51 &   63 &   123 \\
NGC 3516   & &   2.1 &  48 &   6.4 & NP & G  & 0.33 &   43 &   133 \\
NGC 3783   & &   7.2 & 130 &  14.0 & M  & G  & 0.51 &   55 &   108 \\
NGC 4051   & &   1.9 &  39 &   9.0 & NP & G  & 0.21 &   49 &   231 \\
NGC 4593   & &   3.7 &  17 &   5.0 & M  & M  & 0.74 &  218 &   294 \\
NGC 5548   & &   3.2 &  58 &   4.6 & NP & G  & 0.70 &   55 &    79 \\
NGC 7213   & &   3.8 &     &   7.5 & NP & G  & 0.51 &      &       \\
NGC 7469   & &   3.7 &  58 &  23.0 & NP & G  & 0.16 &   64 &   397 \\
\multicolumn{10}{c}{Seyfert 2}\\
\hline
IC 5063    & 370 &  1.0 &    93 &       & S &     &       &   11 &      \\
MCG-5-23-16&  20 &  3.0 &    23 &  15.0 & S & R   &  0.20 &  131 & 652  \\
Mrk 3      & 810 &  1.0 &   347 &   8.1 & S & G   &  0.13 &    3 &   23 \\
Mrk 348    & 133 &  1.3 &    42 &   8.6 & S & G   &  0.15 &   30 &  205 \\
NGC 2110   &  25 &  3.5 &    17 &   5.6 & S & R   &  0.62 &  204 &  329 \\
NGC 2992   &  19 &  1.6 &    91 &   7.3 & S & G   &  0.22 &   18 &   80 \\
NGC 4388   & 380 &  1.4 &    48 &   8.6 & I & R   &  0.16 &   29 &  179 \\
NGC 4507   & 593 &  1.6 &   110 &   6.5 & S & G   &  0.25 &   15 &   59 \\
NGC 526A   &  23 &  1.0 &    27 &       & S &     &       &   38 &      \\
NGC 5506   &  39 &  5.5 &    45 &  20.4 & S & R   &  0.27 &  122 &  453 \\
NGC 7314   &   6 &  2.9 &   6.1 &       & S &     &       &  477 &      \\
\end{tabular}
\caption{X--ray data for Seyfert nuclei with well--determined X--Ray spectra.
$N_H$ is given in $10^{21}$ cm$^{-2}$.
$F_{2-10}$, $F_{5007}$ and $F_{10 \mu m}$ 
are the 2-10 keV, the [OIII]$\lambda$5007  and the N band fluxes
in units of $10^{-11}$, $10^{-14}$ and $10^{-11}$ erg cm $^{-2}$ s${-1}$
respectively. Reference codes for IR data are the same as in Table 1.
The [OIII] fluxes are taken from Whittle 1992. References for  X-ray data
are: NP: Nandra \& Pounds 1994; M: Malaguti et al.\ 1994; S : Smith \&
Done 1996.
}
\label{tabx}
\end{table*}
\nocite{Whi92,NaP94,SmD96,MBC94}

\newpage

\begin{table*}
\begin{tabular*}{13cm}{lccccccc}
 & $\lambda$ & Area & F lim   & Gal & AGN 1 & AGN 2 & $<z_{AGN}>$ \\
 & $\mu$m     &deg$^2$& $\mu$Jy&     &       &       &       \\
ELAIS & 15 & 20 & 1000 & 1300 & 50 & 200 & 0.2\\
CAM T & 15 & 1.5&  200 & 1100 & 40 & 100 & 0.6\\
CAM T & 15 & 0.25&  100 & 360 & 15 & 50 & 0.84 \\
CAM T & 6.7& 0.25&  40  & 600 & 20 & 60 & 0.9 \\
\end{tabular*}

\caption{Predicted  counts of normal galaxies and AGNs for various
ISO surveys.}
\label{tabcounts}
\end{table*}

\end{document}